\documentclass{ws-procs9x6}

\newcommand{\bs}{\boldsymbol}
\newcommand{\acrit}{a_{\rm crit}}
\newcommand{\chr}{$^{52}$Cr}

\newcommand{\mub}[0]{\mu_{\mathrm{B}}}

\newcommand{\add}[0]{a_{\rm dd}}
\newcommand{\edd}[0]{\varepsilon_{\rm dd}}
\newcommand{\udd}[0]{U_{\rm dd}}

\newcommand{\ddi}[0]{dipole-dipole interaction}

\begin{document}

\title{%
A purely dipolar quantum gas
}

\author{T. Lahaye, J. Metz, T. Koch, B. Fr\"ohlich, A. Griesmaier and \underline{T. Pfau}$^*$,}

\address{%
  5. Physikalisches Institut, Universit\"at Stuttgart,
  D-70550 Stuttgart, Germany\\
  $^*$E-mail: t.pfau@physik.uni-stuttgart.de\\
}

\begin{abstract}
We report on experiments exploring the physics of dipolar quantum gases using a {\chr} Bose-Einstein condensate (BEC). By means of a Feshbach resonance, it is possible to reduce the effects of short range interactions and reach a regime where the physics is governed by the long-range, anisotropic {\ddi} between the large ($6\,\mub$) magnetic moments of Chromium atoms. Several dramatic effects of the dipolar interaction are observed: the usual inversion of ellipticity of the condensate during time-of flight is inhibited, the stability of the dipolar gas depends strongly on the trap geometry, and the explosion following the collapse of an unstable dipolar condensate displays $d$-wave like features.
\end{abstract}


\bodymatter

\section{Introduction}

Although quantum gases are very dilute systems, most of their properties are governed by atomic interactions. This allows to use them, for example, as \emph{quantum simulators} to study the many-body physics of systems usually encountered in condensed matter physics~\cite{bloch2008}. However, in all usual quantum gases, the interactions can be described extremely well by a short range, isotropic \emph{contact} potential, whose magnitude is proportional to the $s$-wave scattering length $a$ characterizing low energy collisions.

The {\ddi} taking place between particles having a permanent electric or magnetic dipole moment has radically different properties: it is long-range and anisotropic, as one readily sees on the expression
\begin{equation}
U_{\rm dd}({\bs r})=\frac{\mu_0\mu^2}{4\pi}\,\frac{1-3\cos^2\theta}{r^3}
\label{eq:udd}
\end{equation}
giving the interaction energy $\udd$ between two polarized dipoles separated by ${\bs r}$ ($\theta$ is the angle between ${\bs r}$ and the direction along which the dipoles are pointing). These specific properties have attracted a lot of interest recently, and a large number of theoretical predictions have been made concerning dipolar quantum gases (see e.g. Ref.~\refcite{baranov2008} for a review): for instance, the stability of a dipolar BEC depends crucially on the trap geometry (see section \ref{sec:stab} below); in a quasi two-dimensional trap, the excitation spectrum can display a roton minimum instead of the usual Bogoliubov shape; finally, fascinating new quantum phases (including supersolids) are predicted to occur for dipolar bosons in an optical lattice.

In practice one always has a competition between contact and dipolar interactions; it is therefore useful to define the following (dimensionless) ratio of the dipolar and contact coupling constants:
\begin{equation}
\edd=\frac{\mu_0\mu^2m}{12\pi\hbar^2a}.
\label{eq:edd}
\end{equation}
The numerical factors are chosen in such a way that a \emph{homogeneous} dipolar condensate is unstable against collapse for $\edd>1$. For usual atomic magnetic moments $\mu$ (e.g. for the alkalis), $\edd$ is very small (typically a few $10^{-3}$) and dipolar effects are extremely small. Here, we report on  experiments with {\chr}, which has $\edd\simeq0.16$ due to its large magnetic moment $\mu=6\,\mub$, and which also allows, \emph{via} Feshbach tuning of the scattering length $a$, to even enhance $\edd$.

This paper is organized as follows. We first describe briefly in section~\ref{sec:fr} our experimental setup, with an emphasis on how we use a Feshbach resonance in order to enhance dipolar effects and create a `quantum ferrofluid'. Section~\ref{sec:stab} is devoted to the study of the geometry dependence of the stability of a dipolar BEC. Finally, we describe in section~\ref{sec:collapse} the dynamics following the collapse of an unstable dipolar condensate.

\section{Enhancing dipolar effects using a Feshbach resonance}\label{sec:fr}

\begin{figure}[t]
\begin{center}
\psfig{file=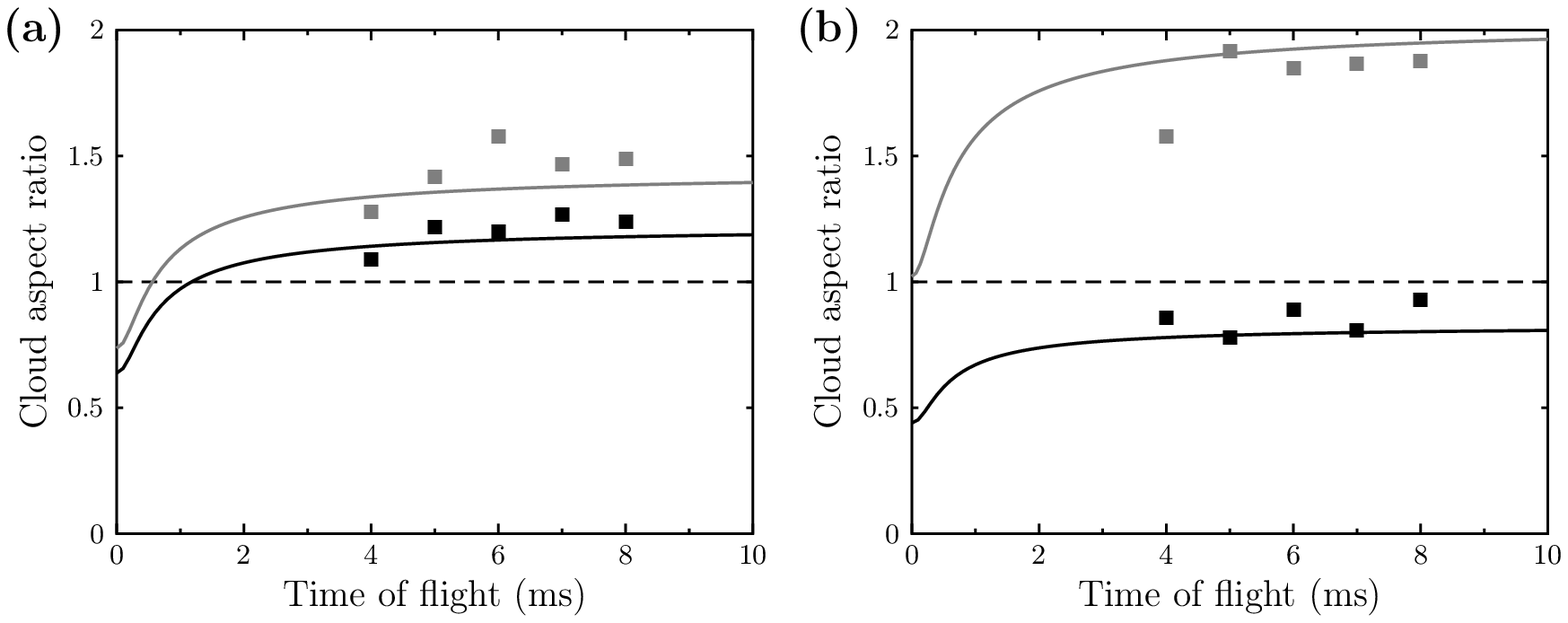,width=4.5in}
\end{center}
\caption{Free expansion of a dipolar condensate for two different orientations of the dipoles with respect to the trap axes. The black (resp. gray) squares and lines correspond to a situation where the dipoles point along the weak (resp. strong) axis of the trap.  {\bf (a)}: Perturbative regime $\edd=0.16$; {\bf (b)}: $\edd=0.75$. In that case, the {\ddi} is strong enough to inhibit the usual inversion of ellipticity in time of flight.}
\label{fig1}
\end{figure}

A BEC of {\chr} containing about 50,000 atoms was obtained in 2005 by evaporative cooling of optically trapped chromium atoms~\cite{griesmaier2005}. Shortly after the achievement of condensation, a first effect of the {\ddi} could be observed in time-of flight experiments~\cite{stuhler2005}: The {\ddi} tends to \emph{elongate} the BEC along the magnetization direction. However, due to the small value of $\edd\simeq0.16$, the dipolar interaction was, in this experiment, only a small perturbation of the contact interaction, which essentially governed the expansion dynamics.

The existence of several Feshbach resonances~\cite{werner2005} in {\chr} opens the possibility to tune the scattering length $a$ using an external magnetic field $B$, according to
\begin{equation*}
a=a_{\rm bg}\left(1-\frac{\Delta}{B-B_0}\right).
\end{equation*}
Here, $a_{\rm bg}\simeq100a_0$ is the $s$-wave scattering length, $B_0$ the resonance position, and $\Delta$ the resonance width.
The broadest Feshbach resonance in {\chr} is located at $B_0=589$~G and has a width $\Delta$ of only $1.5$~G. This implies that the field control at the level of $3\times 10^{-5}$ r.m.s. that we implemented allows us to tune $a$ close to 0 with a resolution of about one Bohr radius.

This `knob' allowing one to change $a$ allowed us to perform time of flight experiments for two different orientations of the dipoles with respect to the trap axes~\cite{lahaye2007}, as in Ref.~\refcite{stuhler2005}, but now for increasing values of $\edd$. Experimental results are shown in Fig.~\ref{fig1}. One clearly sees the dramatic effect of an increase of $\edd$ on the expansion dynamics. In particular, for $\edd\simeq0.75$, the inversion of ellipticity of the condensate during time of flight (the usual `smoking-gun' evidence for BEC) is inhibited by the strong {\ddi}.

The solid lines in Fig.~\ref{fig1} are theoretical predictions (without any adjustable parameters) based on the Gross-Pitaevskii equation (GPE) generalized to take into account the non-local {\ddi} in the description of the macroscopic wavefunction $\psi({\bs r},t)$ of the BEC:
\begin{equation}
i\hbar\frac{\partial \psi}{\partial t}=\left(-\frac{\hbar^2}{2m}\triangle+V_{\rm ext}+g|\psi|^2+\int\! |\psi({\bs r}',t)|^2 \udd({\bs r}-{\bs r}')\,{\rm d}{\bs r}'\right)\psi.
\label{eq:gpe}
\end{equation}
Here $g=4\pi\hbar^2a/m$ is the contact interaction coupling constant.

In this set of experiments, the trap geometry was not very far from spherical, which limited the study of dipolar condensates to values $\edd\lesssim1$. To go beyond this value and reach the purely dipolar regime $\edd\gg1$, we shall now see that one needs to tailor the confining potential, so that the attractive part of the {\ddi} does not destabilize the condensate.

\section{Geometrical stabilization of a purely dipolar condensate}\label{sec:stab}

\subsection{Experimental study}

\begin{figure}[t]
\begin{center}
\psfig{file=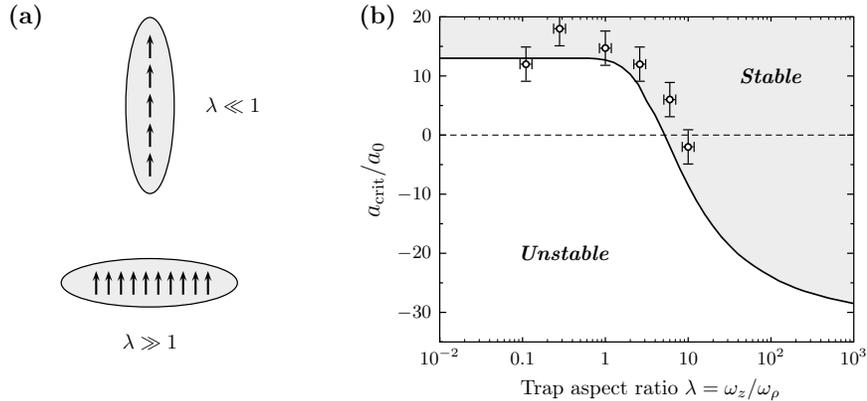,width=4.5in}
\end{center}
\caption{{\bf (a)}: Intuitive picture of the geometry-dependent stability of the dipolar Bose gas. For a prolate trap (aspect ratio $\lambda=\omega_z/\omega_\rho$ smaller than one) with the dipoles pointing along the weak axis of the trap, the {\ddi} is essentially attractive; such a condensate is thus unstable. For an oblate trap, the {\ddi} is essentially repulsive and the BEC is stable. {\bf (b)}: Stability diagram of a dipolar BEC in the plane $(\lambda,a)$. The points with error bars are the experimental results for the critical scattering length $\acrit$ below which no condensate is observed; the solid line is the stability threshold obtained with a simple gaussian ansatz (see text).}
\label{fig2}
\end{figure}

It is well known that a BEC with attractive contact interactions is unstable against long-wavelength fluctuations (this \emph{phonon instability} leads to a collapse of the BEC having $a<0$). As the {\ddi} has an attractive part for dipoles in a `head-to-tail' configuration --- see equation (\ref{eq:udd}) for $\theta\simeq0$ ---, it is intuitively clear that in a prolate trap with the dipoles pointing along the weak direction of the trap [see Fig.~\ref{fig2}{\bf(a)}], the net effect of the dipolar interaction is attractive. Thus, in this configuration, one expects that when $a$ is reduced, the condensate becomes unstable, at a critical value $\acrit$ which should be positive (the small repulsive contact interaction being unable, at this point, to counteract the dipolar attraction). Conversely, in an oblate trap with the dipoles pointing along the strong confinement direction, the critical scattering length should be negative, and a \emph{purely dipolar quantum gas} can be stabilized.

In Ref.~\refcite{koch2008}, this geometry-dependent stability of a dipolar condensate was studied experimentally. A long period ($\simeq8$~$\mu$m) optical lattice, obtained by interfering two laser beams at 1064~nm under a small angle of $8^\circ$, was superimposed onto the optical dipole trap, allowing us to realize traps with cylindrical symmetry around the $z$-axis (polarization direction) and having an aspect ratio $\lambda\equiv\omega_z/\omega_\rho$ that could be varied over two orders of magnitude (from $\sim0.1$ to $\sim 10$) while keeping the average trapping frequency $\bar{\omega}=(\omega_z\omega_\rho^2)^{1/3}$ constant. The experiment consists in creating a BEC in a trap with a given aspect ratio $\lambda$, then ramping $a$ to a final value $a_{\rm f}$, and finally measuring the atom number $N$ in the condensate. One observes that when $a_{\rm f}$ is decreased below a critical value $\acrit$, $N$ suddenly drops to zero. We stress that for all the traps we used, the condensate density was roughly the same.

Figure~\ref{fig2}{\bf(b)} shows the measured $\acrit$ as a function of $\lambda$ and clearly displays the expected behavior: for small $\lambda$ (prolate traps), $\acrit$ is positive, and starts to decrease when the trap becomes more oblate. For $\lambda\simeq10$, one has $\acrit\simeq0$, meaning that a purely dipolar quantum gas ($\edd\to\infty$) can be stabilized by an appropriate trap geometry.

\subsection{A simple theoretical model}

A simple way to go beyond the qualitative picture above and obtain an estimate for the instability threshold $\acrit(\lambda)$ is to use a variational method. Inserting a Gaussian ansatz (with the axial and radial sizes $\sigma_z$ and $\sigma_\rho$ as variational parameters) into the Gross-Pitaevskii energy functional whose minimization gives the GPE (\ref{eq:gpe}), one obtains the following energy to minimize:
\begin{eqnarray}
E(\sigma_\rho,\sigma_z)&=&
\frac{N\hbar\bar{\omega}}{4}\left(\frac{2}{\sigma_r^2}+\frac{1}{\sigma_z^2}\right)
+\frac{N\hbar\bar{\omega}}{4\lambda^{2/3}}\left(2\sigma_r^2+\lambda^2\sigma_z^2\right)\nonumber\\
&&+\frac{N^2\hbar\bar{\omega}a}{\sqrt{2\pi}\ell}\,\frac{1}{\sigma_\rho^2\sigma_z}\left[1-\edd f\left(\frac{\sigma_\rho}{\sigma_z}\right)\right],
\label{eq:ansatz}
\end{eqnarray}
where $\ell=\sqrt{\hbar/(m\bar{\omega})}$. The first two terms are the kinetic and potential energies, while the third arises from contact and dipolar interactions. The function $f$ is monotonically decreasing from 1 to $-2$ as a result of the anisotropy of the dipolar interaction. For a given $\lambda$, one can find a (possibly local) minimum of $E$ at finite values of $(\sigma_\rho,\sigma_z)$ if and only if $a$ is larger than a critical value: this defines the stability threshold $\acrit(\lambda)$ within this model.

The solid line in Fig.~\ref{fig2}{\bf(b)} is the result obtained with this simple procedure, for the experimental parameters $\bar{\omega}=2\pi\times 800$~Hz and $N\simeq2\times10^4$.  One obtains a relatively good agreement with experimental data. A numerical solution of the GPE (\ref{eq:gpe}) gives even better agreement with measurements~\cite{bohn2008}.

Equation (\ref{eq:ansatz}) allows one to understand easily, in the $N\to\infty$ limit, the behavior of $\acrit(\lambda)$ for $\lambda\to0$ and $\lambda\to\infty$. Indeed, in this limit, it is the sign of the interaction term which determines the stability; therefore one has
\begin{equation}
\left\{
\begin{array}{rcl}
\lambda\to 0&:& {\rm BEC\; unstable\; if\;} a<\add\\
\lambda\to \infty&:& {\rm BEC\; unstable\; if\;} a<-2\add,
\end{array}
\right.
\label{eq:limits}
\end{equation}
where $\add$ is the length defined in such a way that $\edd=\add/a$ (for {\chr}, one readily calculates, with help of (\ref{eq:edd}), that $\add\simeq15a_0$). It is apparent on Fig.~\ref{fig2}{\bf(b)} that for $N=2\times10^4$, the results are already close to the $N\to\infty$ limit (\ref{eq:limits}).

\section{$d$-wave collapse of a dipolar condensate}\label{sec:collapse}

It is natural to ask what happens if one drives the condensate into the unstable regime, e.g. by decreasing the scattering length below $\acrit$. In the case of pure contact interactions, a collapse of the condensate, followed by an explosion of a `remnant' BEC (\emph{Bose-Nova}), has been observed in several systems~\cite{sackett1999,gerton2000,roberts2001,donley2001}. More recently, the formation of soliton trains has also been reported~\cite{strecker2002,cornish2006}.

\begin{figure}[t]
\begin{center}
\psfig{file=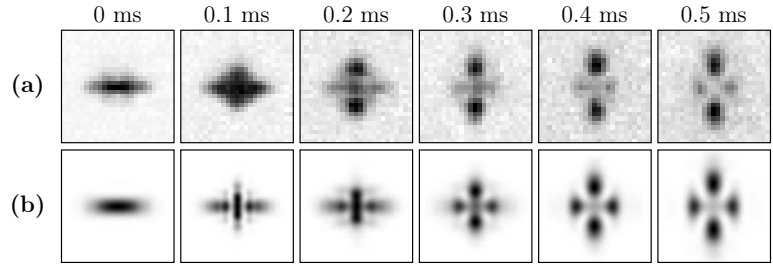,width=4in}
\end{center}
\caption{{\bf(a)}: Experimental images of the `exploding' remnant condensate after the collapse, as a function of the holding time. {\bf(b)}: Result of a numerical simulation of the experiment, without any adjustable parameter. The field of view is $130\,\mu$m $\times$ $130\,\mu$m.}
\label{fig3}
\end{figure}

We have studied the collapse dynamics of a dipolar condensate~\cite{lahaye2008} (in a roughly spherical trap) by ramping down rapidly the scattering length to a final value of $\sim5a_0<\acrit$, then waiting an adjustable holding time, and performing a time of flight of 8~ms before imaging the cloud. Figure~\ref{fig3}{\bf(a)} presents the evolution of the condensate when the holding time is varied. One observes that the cloud, initially elongated along the magnetization direction $z$ (horizontal axis on the figure) acquires rapidly a complicated structure with a four-fold symmetry, corresponding to a density distribution having a torus-like component close to the plane $z=0$ and two `blobs' close to the $z$-axis. Interestingly, this angular symmetry of the cloud is very close to the one of a $d$-wave $\propto(1-3\cos^2\theta)$, i.e. precisely the symmetry of the {\ddi} (\ref{eq:udd}). During the same time period, the atom number in the condensate strongly decreases due to the three-body losses occurring because of the high densities transiently reached during the collapse.

The group of M. Ueda in Tokyo performed a three dimensional numerical simulation of the GPE (\ref{eq:gpe}), in which all input parameters were given their experimentally measured value~\cite{lahaye2008}. Three-body losses were accounted for by adding the imaginary term
\begin{equation*}
i\hbar\left.\frac{\partial\psi}{\partial t}\right|_{3\rm\,body}=-\frac{i\hbar L_3}{2}|\psi|^4\psi
\end{equation*}
to Eqn. (\ref{eq:gpe}), where $L_3\simeq2\times10^{-40}$~m$^6/$s is the measured three-body loss coefficient. Figure~\ref{fig3}{\bf(b)} represents the results of the simulation. The agreement is excellent, all the more if one keeps in mind that no adjustable parameter is introduced. Let's mention that the inclusion of a small delay (also measured independently) in the time variation $a(t)$ of the scattering length, due to eddy currents in the vacuum chamber, had to be included to achieve a quantitative agreement! The simulation also reproduces quantitatively the time dependence of the condensate atom number.

A fascinating prediction of the numerical simulation is the spontaneous formation, during the collapse, of two quantized vortex rings with opposite circulation (and charge $\pm1$), as a result of the strongly anisotropic collapse: the collapse in the radial directions is fast and quickly followed by an outward flow, while axially the flow is still inward, thus giving rise to the circulation. Detecting experimentally the presence of vortex rings is very challenging, but might be done by using interferometric techniques (e.g. matter wave heterodyning) to reveal the winding of the phase of the BEC wavefunction around the topological defects.

\section{Outlook}

The results presented in this paper are the first dramatic manifestations of dipolar effects in quantum gases, and pave the way for future studies involving even more strongly interacting dipolar systems, especially the ones that may be obtained using the permanent electric dipole moments of heteronuclear molecules in their ground state. Due to the large value of such dipole moments (on the order of one Debye), the long-range character of the dipolar interaction could then be used to achieve novel quantum phases in optical lattices~\cite{baranov2008}, as well as to implement promising quantum information processing schemes~\cite{demille2002}.

However, already in the case of the comparatively weaker magnetic dipoles, extremely interesting theoretical proposals deserve experimental study; to mention only one example, the generation of two-dimensional solitons~\cite{pedri2005,tikhonenkov2008} (whose stability arises from the long-range character of the {\ddi}) is a very appealing experiment.

\section*{Acknowledgments}

We thank H. Saito, Y. Kawaguchi and M. Ueda for their collaboration on Ref.~\refcite{lahaye2008}. We acknowledge support by the German Science Foundation (SFB/TRR 21, SPP 1116), the Landesstiftung Baden-W\"urttenberg and the EU (Marie-Curie Grant MEIF-CT-2006-038959 to T. L.).

\end{document}